\pdfoutput=1
\documentclass[12pt,a4paper]{article}

\usepackage{ifthen} 
\newboolean{pdflatex}
\setboolean{pdflatex}{true} 

\newboolean{articletitles}
\setboolean{articletitles}{true} 

\newboolean{uprightparticles}
\setboolean{uprightparticles}{false} 

\def\paperauthors{LHCb collaboration} 
\def\paperasciititle{Measurement of the photon polarization in Lb to Lambda gamma decays} 
\def\papertitle{Measurement of the photon polarization in \LbLzGam decays} 
\def\paperkeywords{{High Energy Physics}, {LHCb}} 
\def\papercopyright{\the\year\ CERN for the benefit of the LHCb collaboration} 
\def\paperlicence{CC BY 4.0 licence}
\def\paperlicenceurl{https://creativecommons.org/licenses/by/4.0/}

\usepackage[top=1in, bottom=1.25in, left=1in, right=1in]{geometry}

\usepackage{microtype}
\usepackage{lineno}  
\usepackage{xspace} 
\usepackage{caption} 

\usepackage{graphicx}  
\usepackage{color}
\usepackage{colortbl}
\graphicspath{{./figs/}} 

\usepackage{amsmath} 
\usepackage{amssymb}
\usepackage{amsfonts}
\usepackage{upgreek} 

\newcommand*\patchAmsMathEnvironmentForLineno[1]{%
\expandafter\let\csname old#1\expandafter\endcsname\csname #1\endcsname
\expandafter\let\csname oldend#1\expandafter\endcsname\csname
end#1\endcsname
 \renewenvironment{#1}%
   {\linenomath\csname old#1\endcsname}%
   {\csname oldend#1\endcsname\endlinenomath}%
}
\newcommand*\patchBothAmsMathEnvironmentsForLineno[1]{%
  \patchAmsMathEnvironmentForLineno{#1}%
  \patchAmsMathEnvironmentForLineno{#1*}%
}
\AtBeginDocument{%
\patchBothAmsMathEnvironmentsForLineno{equation}%
\patchBothAmsMathEnvironmentsForLineno{align}%
\patchBothAmsMathEnvironmentsForLineno{flalign}%
\patchBothAmsMathEnvironmentsForLineno{alignat}%
\patchBothAmsMathEnvironmentsForLineno{gather}%
\patchBothAmsMathEnvironmentsForLineno{multline}%
\patchBothAmsMathEnvironmentsForLineno{eqnarray}%
}

\usepackage{hyperxmp}

\usepackage[pdftex,
            pdfauthor={\paperauthors},
            pdftitle={\paperasciititle},
            pdfkeywords={\paperkeywords},
            pdfcopyright={Copyright (C) \papercopyright},
            pdflicenseurl={\paperlicenceurl}]{hyperref}

\usepackage[colorinlistoftodos,textsize=scriptsize]{todonotes}

\usepackage[bottom,flushmargin,hang,multiple]{footmisc}

\usepackage[all]{hypcap} 

\usepackage{xspace} 
\usepackage{upgreek}

\def\lhcb   {\mbox{LHCb}\xspace}

\def\babar  {\mbox{BaBar}\xspace}
\def\belle  {\mbox{Belle}\xspace}

\def\MagUp {\mbox{\em Mag\kern -0.05em Up}\xspace}

\ifthenelse{\boolean{uprightparticles}}%
{
 
 \def\Pgamma      {\ensuremath{\upgamma}\xspace}

 \def\Peta        {\ensuremath{\upeta}\xspace}

 \def\Ppi         {\ensuremath{\uppi}\xspace}

 \def\Ppsi        {\ensuremath{\uppsi}\xspace}

 \def\PDelta      {\ensuremath{\Delta}\xspace}                 
 \def\PXi         {\ensuremath{\Xi}\xspace}                 
 \def\PLambda     {\ensuremath{\Lambda}\xspace}                 
 \def\PSigma      {\ensuremath{\Sigma}\xspace}                 
 \def\POmega      {\ensuremath{\Omega}\xspace}                 
 \def\PUpsilon    {\ensuremath{\Upsilon}\xspace}

 \def\PB      {\ensuremath{\mathrm{B}}\xspace}                 
                  
 \def\PD      {\ensuremath{\mathrm{D}}\xspace}

 \def\PJ      {\ensuremath{\mathrm{J}}\xspace}                 
 \def\PK      {\ensuremath{\mathrm{K}}\xspace}

 \def\Pb      {\ensuremath{\mathrm{b}}\xspace}                 
 \def\Pc      {\ensuremath{\mathrm{c}}\xspace}                 
                  
 \def\Pe      {\ensuremath{\mathrm{e}}\xspace}

 \def\Pi      {\ensuremath{\mathrm{i}}\xspace}

 \def\Pp      {\ensuremath{\mathrm{p}}\xspace}

 \def\Ps      {\ensuremath{\mathrm{s}}\xspace}

 \def\thebaroffset{0.0em}
}
{
 
 \def\Pgamma      {\ensuremath{\gamma}\xspace}

 \def\Peta        {\ensuremath{\eta}\xspace}

 \def\Ppi         {\ensuremath{\pi}\xspace}

 \def\Ppsi        {\ensuremath{\psi}\xspace}                 
                  
 \mathchardef\PDelta="7101
 \mathchardef\PXi="7104
 \mathchardef\PLambda="7103
 \mathchardef\PSigma="7106
 \mathchardef\POmega="710A
 \mathchardef\PUpsilon="7107
                  
 \def\PB      {\ensuremath{B}\xspace}                 
                  
 \def\PD      {\ensuremath{D}\xspace}

 \def\PJ      {\ensuremath{J}\xspace}                 
 \def\PK      {\ensuremath{K}\xspace}

 \def\Pb      {\ensuremath{b}\xspace}                 
 \def\Pc      {\ensuremath{c}\xspace}                 
                  
 \def\Pe      {\ensuremath{e}\xspace}

 \def\Pi      {\ensuremath{i}\xspace}

 \def\Pp      {\ensuremath{p}\xspace}

 \def\Ps      {\ensuremath{s}\xspace}

 \def\thebaroffset{0.18em}
}
\newcommand{\offsetoverline}[2][\thebaroffset]{\kern #1\overline{\kern -#1 #2}}%

\makeatletter
\ifcase \@ptsize \relax
  \newcommand{\miniscule}{\@setfontsize\miniscule{4}{5}}
\or
  \newcommand{\miniscule}{\@setfontsize\miniscule{5}{6}}
\or
  \newcommand{\miniscule}{\@setfontsize\miniscule{5}{6}}
\fi
\makeatother

\DeclareRobustCommand{\optbar}[1]{\shortstack{{\miniscule (\rule[.5ex]{1.25em}{.18mm})}
  \\ [-.7ex] $#1$}}

\def\epem       {{\ensuremath{\Pe^+\Pe^-}}\xspace}

\def\g      {{\ensuremath{\Pgamma}}\xspace}

\def\squark    {{\ensuremath{\Ps}}\xspace}

\def\cquark    {{\ensuremath{\Pc}}\xspace}

\def\bquark    {{\ensuremath{\Pb}}\xspace}
\def\bquarkbar {{\ensuremath{\overline \bquark}}\xspace}

\def\pion   {{\ensuremath{\Ppi}}\xspace}
\def\piz    {{\ensuremath{\pion^0}}\xspace}

\def\pim    {{\ensuremath{\pion^-}}\xspace}

\def\kaon    {{\ensuremath{\PK}}\xspace}

\def\KorKbar {\kern \thebaroffset\optbar{\kern -\thebaroffset \PK}{}\xspace}

\def\Km      {{\ensuremath{\kaon^-}}\xspace}

\def\KS      {{\ensuremath{\kaon^0_{\mathrm{S}}}}\xspace}

\def\Kstarz  {{\ensuremath{\kaon^{*0}}}\xspace}

\newcommand{\etaz}{\ensuremath{\Peta}\xspace}

\def\D       {{\ensuremath{\PD}}\xspace}

\def\DorDbar {\kern \thebaroffset\optbar{\kern -\thebaroffset \PD}\xspace}

\def\Dp      {{\ensuremath{\D^+}}\xspace}
\def\Dm      {{\ensuremath{\D^-}}\xspace}

\def\DpDm    {\ensuremath{\Dp {\kern -0.16em \Dm}}\xspace}

\def\B       {{\ensuremath{\PB}}\xspace}

\def\BorBbar {\kern \thebaroffset\optbar{\kern -\thebaroffset \PB}\xspace}

\def\Bd      {{\ensuremath{\B^0}}\xspace}

\def\BdorBdbar {\kern \thebaroffset\optbar{\kern -\thebaroffset \Bd}\xspace}

\def\Bs      {{\ensuremath{\B^0_\squark}}\xspace}

\def\BsorBsbar {\kern \thebaroffset\optbar{\kern -\thebaroffset \Bs}\xspace}

\def\jpsi     {{\ensuremath{{\PJ\mskip -3mu/\mskip -2mu\Ppsi}}}\xspace}

\def\Y#1S{\ensuremath{\PUpsilon{(#1S)}}\xspace}

\def\proton      {{\ensuremath{\Pp}}\xspace}

\def\Lz          {{\ensuremath{\PLambda}}\xspace}
\def\Lbar        {{\ensuremath{\offsetoverline{\PLambda}}}\xspace}
\def\LorLbar     {\kern \thebaroffset\optbar{\kern -\thebaroffset \PLambda}\xspace}

\def\Lb           {{\ensuremath{\Lz^0_\bquark}}\xspace}
\def\Lbbar        {{\ensuremath{\Lbar{}^0_\bquark}}\xspace}

\def\BF         {{\ensuremath{\mathcal{B}}}\xspace}

\newcommand{\decay}[2]{\ensuremath{#1\!\to #2}\xspace} 

\def\to                 {\ensuremath{\rightarrow}\xspace}

\def\CP                {{\ensuremath{C\!P}}\xspace}

\def\BsPhiGam     {\decay{\Bs}{\phi \g}}
\def\BdKstGam     {\decay{\Bd}{\Kstarz \g}}

\def\BdKstee  {\decay{\Bd}{\Kstarz\epem}}

\def\AT#1     {\ensuremath{A_{\mathrm{T}}^{#1}}\xspace}           
\def\btosgam  {\decay{\bquark}{\squark \g}}

\def\C#1      {\ensuremath{\mathcal{C}_{#1}}\xspace}                       
\def\Cp#1     {\ensuremath{\mathcal{C}_{#1}^{'}}\xspace}                    
\def\Ceff#1   {\ensuremath{\mathcal{C}_{#1}^{\mathrm{(eff)}}}\xspace}        
\def\Cpeff#1  {\ensuremath{\mathcal{C}_{#1}^{'\mathrm{(eff)}}}\xspace}       
\def\Ope#1    {\ensuremath{\mathcal{O}_{#1}}\xspace}                       
\def\Opep#1   {\ensuremath{\mathcal{O}_{#1}^{'}}\xspace}                    

\newcommand{\aunit}[1]{\ensuremath{\text{\,#1}}}       

\newcommand{\tev}{\aunit{Te\kern -0.1em V}\xspace}
\newcommand{\gev}{\aunit{Ge\kern -0.1em V}\xspace}
\newcommand{\mev}{\aunit{Me\kern -0.1em V}\xspace}
\newcommand{\kev}{\aunit{ke\kern -0.1em V}\xspace}
\newcommand{\ev}{\aunit{e\kern -0.1em V}\xspace}
 
\newcommand{\mevc}{\ensuremath{\aunit{Me\kern -0.1em V\!/}c}\xspace}
\newcommand{\gevc}{\ensuremath{\aunit{Ge\kern -0.1em V\!/}c}\xspace}
\newcommand{\mevcc}{\ensuremath{\aunit{Me\kern -0.1em V\!/}c^2}\xspace}
\newcommand{\gevcc}{\ensuremath{\aunit{Ge\kern -0.1em V\!/}c^2}\xspace}

\def\mm   {\aunit{mm}\xspace}

\def\fb   {\ensuremath{\aunit{fb}}\xspace}
\def\invfb   {\ensuremath{\fb^{-1}}\xspace}

\def\gsim{{~\raise.15em\hbox{$>$}\kern-.85em
          \lower.35em\hbox{$\sim$}~}\xspace}
\def\lsim{{~\raise.15em\hbox{$<$}\kern-.85em
          \lower.35em\hbox{$\sim$}~}\xspace}

\def\pt         {\ensuremath{p_{\mathrm{T}}}\xspace}

\def\et         {\ensuremath{E_{\mathrm{T}}}\xspace}

\def\evtgen     {\mbox{\textsc{EvtGen}}\xspace}

\def\geant      {\mbox{\textsc{Geant4}}\xspace}

\def\photos     {\mbox{\textsc{Photos}}\xspace}

\def\pythia     {\mbox{\textsc{Pythia}}\xspace}

\def\tell1  {TELL1\xspace}
\def\ukl1   {UKL1\xspace}

\newcommand{\eg}{\mbox{\itshape e.g.}\xspace}

\def\btosgam  {\decay{\bquark}{\squark\g}}

\def\LbLzGam  {\decay{\Lb}{\Lz \g}}

\def\Lzppi {\decay{\Lz}{\proton\pim}}

\def\LbLzJpsi {\decay{\Lb}{\Lz\jpsi}}

\def\LbpKJpsi {\decay{\Lb}{\proton\Km\jpsi}}
\def\LbLzEta {\decay{\Lb}{\Lz \etaz}}
\def\Etagg	{\decay{\etaz}{\g\g}}

\def\LbLzJpsi		{\decay{\Lb}{\Lz\jpsi}}

\def\Jpsimumu  {\decay{\jpsi}{\mu^+\mu^-}}

\def\ag {\ensuremath{\alpha_{\g}}\xspace}
\def\aL {\ensuremath{\alpha_{\Lz}}\xspace}

\def\LbbarLzbarGam  {\decay{\Lbbar}{\Lbar \g}}
\def\agp {\ensuremath{\alpha_{\g}^{+}}\xspace}
\def\agm {\ensuremath{\alpha_{\g}^{-}}\xspace}

\def\aLp {\ensuremath{\alpha_{\Lz}^{+}}\xspace}
\def\aLm {\ensuremath{\alpha_{\Lz}^{-}}\xspace}

\def\tp {\ensuremath{\theta_p}\xspace}
\def\ctp {\ensuremath{\cos{\tp}}\xspace}

\usepackage{cite} 
\usepackage{mciteplus}

\usepackage{longtable} 

\begin{document}

\renewcommand{\thefootnote}{\fnsymbol{footnote}}
\setcounter{footnote}{1}


\begin{titlepage}
\pagenumbering{roman}

\vspace*{-1.5cm}
\centerline{\large EUROPEAN ORGANIZATION FOR NUCLEAR RESEARCH (CERN)}
\vspace*{1.5cm}
\noindent
\begin{tabular*}{\linewidth}{lc@{\extracolsep{\fill}}r@{\extracolsep{0pt}}}
\ifthenelse{\boolean{pdflatex}}
{\vspace*{-1.5cm}\mbox{\!\!\!\includegraphics[width=.14\textwidth]{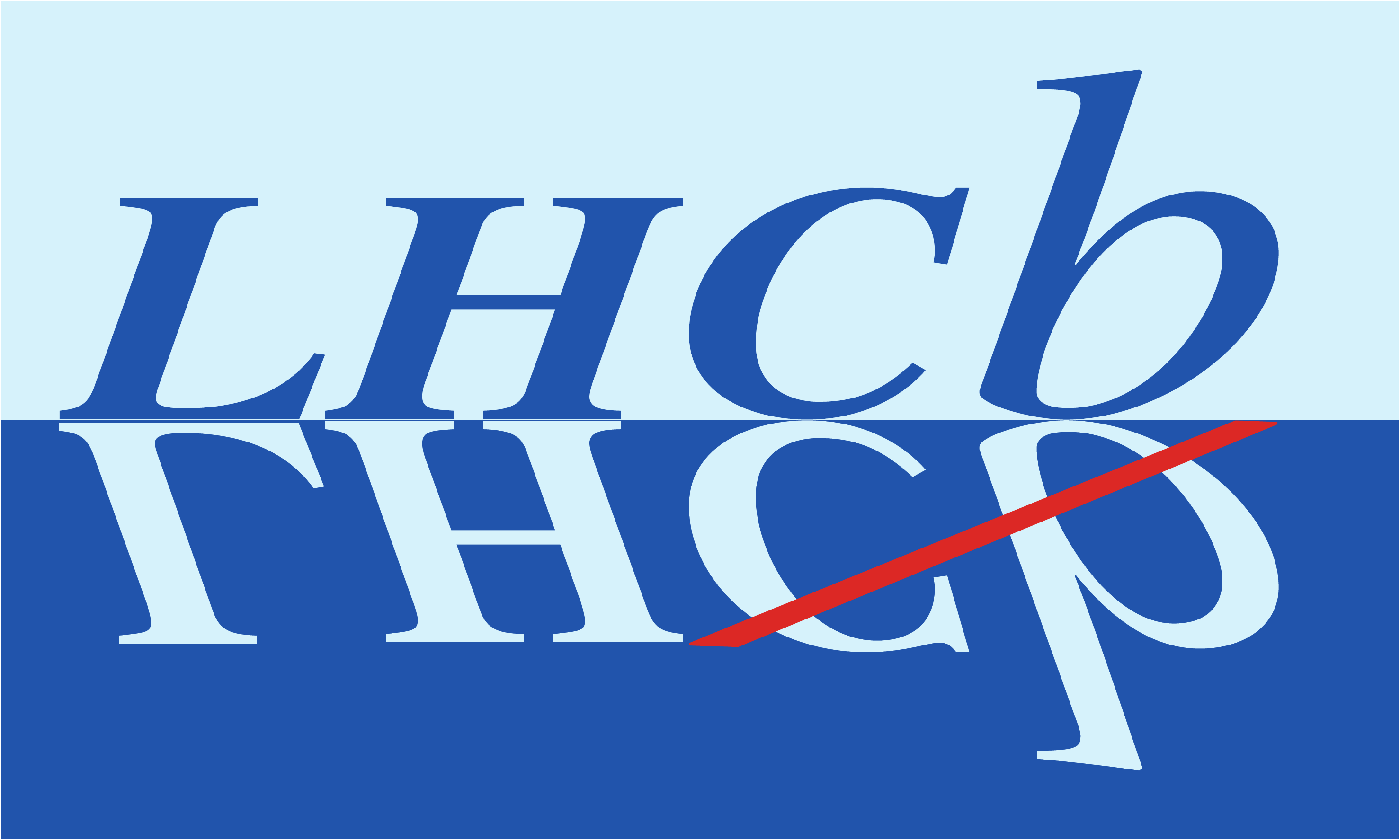}} & &}%
{\vspace*{-1.2cm}\mbox{\!\!\!\includegraphics[width=.12\textwidth]{figs/lhcb-logo.eps}} & &}%
\\
 & & CERN-EP-2021-230 \\  
 & & LHCb-PAPER-2021-030 \\  
 & & 15 April 2022 \\ 
 & & \\
\end{tabular*}

\vspace*{4.0cm}

{\normalfont\bfseries\boldmath\huge
\begin{center}
  \papertitle 
\end{center}
}

\vspace*{2.0cm}

\begin{center}
\paperauthors\footnote{Authors are listed at the end of this paper.}
\end{center}

\vspace{\fill}

\begin{abstract}
  \noindent
  The photon polarization in \btosgam transitions is measured for the first time in radiative \bquark-baryon decays exploiting the unique spin structure of \LbLzGam decays. A data sample corresponding to an integrated luminosity of $6\invfb$ collected by the \lhcb experiment in $pp$ collisions at a center-of-mass energy of $13\tev$ is used. 
  The photon polarization is measured to be $\ag= 0.82^{\,+\,0.17\,+\,0.04}_{\,-\,0.26\,-\,0.13}$, where the first uncertainty is statistical and the second systematic. This result is in agreement with the Standard Model prediction and previous measurements in \bquark-meson decays.
  Charge-parity breaking effects are studied for the first time in this observable and found to be consistent with \CP symmetry.

\end{abstract}

\vspace*{2.0cm}

\begin{center}
 Published in Phys. Rev. D105 (2022) L051104

\end{center}

\vspace{\fill}

{\footnotesize 
\centerline{\copyright~\papercopyright. \href{\paperlicenceurl}{\paperlicence}.}}
\vspace*{2mm}

\end{titlepage}


\newpage
\setcounter{page}{2}
\mbox{~}
%
%
%
%


\renewcommand{\thefootnote}{\arabic{footnote}}
\setcounter{footnote}{0}

\cleardoublepage


\pagestyle{plain} 
\setcounter{page}{1}
\pagenumbering{arabic}


Rare decays of hadrons containing \bquark-quarks mediated by flavor-changing neutral currents (FCNC) are suppressed in the Standard Model (SM) since they occur only at higher order in the perturbative expansion. They are sensitive to physics beyond the SM (BSM) which can give additional contributions and alter the decay properties. 
Radiative decays of the type \btosgam are FCNC, mediated by electroweak-loop transitions which produce a final state photon. 
In the SM, due to the chirality of the electroweak interaction, the emitted photons are polarized.
The photon polarization is defined as the normalized difference between the number of left-handed ($\gamma_L$) and right-handed ($\gamma_R$) photons produced in radiative decays,
\begin{equation}
    \alpha_\gamma \equiv \frac{\gamma_L -\gamma_R}{\gamma_L +\gamma_R},
    \label{eq:photon_polarization}
\end{equation}
with photons emitted in \bquark decays being predominantly left-handed. 
Right-handed contributions arise only due to chirality flips in the outgoing \squark-quark line of the Feynman diagram, which are suppressed by the ratio of the \squark and \bquark quark masses.
At leading order in the SM, the photon polarization is predicted to be $\alpha_\gamma= (1-|r|^2)/(1+|r|^2)$, with $|r| \approx m_{\squark}/m_{\bquark}$~\cite{Atwood:1997zr,Hiller:2001zj,Legger:2006cq}. Effects arising at next-to-leading order are estimated to be at the percent level~\cite{Hiller:2001zj}.
Therefore, a larger right-handed polarization, which would lower the value of \ag, could be a clear indication of BSM physics~\cite{Atwood:1997zr,Kou:2013gna}.

The photon polarization in \btosgam transitions has been indirectly probed by the \babar, \belle and \lhcb experiments through the measurement of mixing-induced charge-parity (\CP) asymmetries in \Bd and \Bs decays~\cite{BaBar:2008okc,Belle:2006pxp,LHCb-PAPER-2019-015} and through the angular analysis of \BdKstee decays at very low dielectron mass~\cite{LHCb-PAPER-2020-020}, with the latter providing the strongest constraints on right-handed currents.
Radiative decays of \bquark baryons offer a unique opportunity for a direct measurement of the photon polarization due to the nonzero spin of the initial- and final-state particles~\cite{Gremm:1995nx}. The \LbLzGam decay\footnote{The inclusion of charge-conjugate processes is implied
    throughout, except in the discussion of asymmetries.} is of special interest due to the weak decay of the \Lz baryon, which probes the helicity structure of the \btosgam transition through the measurement of the \Lz helicity~\cite{Mannel:1997xy,Hiller:2001zj}.
The \LbLzGam decay was first observed in data recorded by \lhcb during 2016~\cite{LHCb-PAPER-2019-010}.
The photon-polarization sensitivity using this decay was estimated in Ref.~\cite{AngDists}.

The angular distribution of \LbLzGam decays, where the \Lz baryon decays into a proton and a pion, is given by the differential width~\cite{Legger:2006cq} 
\begin{equation}
    \label{eq:Lb_AngDist}
    \frac{\text{d}\Gamma}{\text{d}\left(\cos{\theta_p}\right)} \propto  1-\alpha_{\gamma}\alpha_{\Lambda} \cos{\theta_p},
\end{equation}
where \aL is the $\Lz$ weak decay parameter, which describes the interference between the parity-violating s-wave and parity-conserving p-wave of the \Lzppi decay~\cite{Lee:1957qs}, and \tp is the angle between the proton momentum and the negative \Lb momentum in the \Lz rest frame. The photon direction is integrated over in this study since the distribution in this variable is flat for unpolarized \Lb baryons, as produced at the LHC~\cite{CMS:2018wjk,LHCb-PAPER-2020-005}.

Contrary to \bquark decays, photons produced in \bquarkbar decays are expected to be predominantly right-handed in the SM, resulting in a photon polarization of $\alpha_\gamma= -1$.
A discrepancy in the absolute value of the photon polarisation in \bquark and \bquarkbar decays would be a hint of \CP asymmetry in these transitions.
In the SM, \CP asymmetries in \btosgam decays are estimated to be less than $\mathcal{O}(1\%)$~\cite{Hiller:2001zj}. However, new \CP-violating BSM contributions can produce asymmetries of $\mathcal{O}(10\%)$~\cite{Hiller:2001zj}. Experimentally, the strongest constraint on direct \CP violation in \btosgam transitions is from \BdKstGam decays~\cite{LHCb-PAPER-2012-019}. Measurements of angular asymmetries and \CP-conjugated decay branching ratios can separate the \CP-odd and \CP-even components of both the SM and opposite chirality contributions to the \LbLzGam decay rate~\cite{Hiller:2001zj}. 

This Letter reports the first measurement of the photon polarization in \LbLzGam decays, with \Lzppi, using a data sample corresponding to an integrated luminosity of $6\invfb$ collected by the \lhcb experiment in proton-proton ($pp$) collisions at a center-of-mass energy of $13 \tev$ during 2015--2018.
Additionally, the first study of \CP angular asymmetries in \btosgam decays is performed.

The \lhcb detector~\cite{LHCb-DP-2008-001,LHCb-DP-2014-002} is a single-arm forward
spectrometer covering the \mbox{pseudorapidity} range $2<\eta <5$,
designed for the study of particles containing \bquark or \cquark
quarks. 
The detector elements that are relevant for this analysis are: a silicon-strip vertex detector surrounding the $pp$ interaction
region that allows \cquark\ and \bquark\ hadrons to be identified from their characteristically long
flight distance; a tracking system that provides a measurement of the momentum of charged
particles; two ring-imaging Cherenkov detectors that are able to discriminate between
different species of charged hadrons; a calorimeter system consisting of scintillating-pad and preshower detectors, an electromagnetic calorimeter (ECAL) and a hadronic calorimeter, that enables the reconstruction and identification of photons, electrons and hadrons; and a muon system composed of alternating layers of iron and multiwire proportional chambers.

Charged and neutral clusters in the ECAL are separated by extrapolating reconstructed tracks to the ECAL; photons and neutral pions are distinguished by their cluster shape and energy distribution. For decays with
high-energy photons in the final state, \eg \LbLzGam, a mass resolution around $100 \mev$ is achieved\footnote{Natural units
      with $\hbar=c=1$ are used throughout.}~\cite{LHCb-PAPER-2016-034, LHCb-PAPER-2011-042}, dominated by the photon energy resolution.
The online event selection is performed by a trigger~\cite{LHCb-DP-2012-004}, 
consisting of a hardware stage followed by a software stage which fully reconstructs the event.
The hardware trigger requires events to have an ECAL cluster with an energy component transverse to the beam, \et, above a threshold varying between $2.1 - 3.0\gev$. The software trigger first requires at least one charged particle to have
transverse momentum $\pt > 1 \gev$ and to be inconsistent with originating from any primary $pp$ collision vertex (PV).
Subsequently, \Lz candidates are formed from two tracks significantly displaced from any PV, and 
combined with a high-\et photon to identify decays consistent with the signal mode. In the offline selection, trigger signals are associated with
reconstructed particles and only events where the trigger was activated by the decay products of the signal candidate are kept.


Simulated events are used to model the effects of the 
detector geometry and the selection requirements. 
In the simulation, $pp$ collisions are generated using \pythia~\cite{Sjostrand:2007gs,*Sjostrand:2006za} with a specific \lhcb configuration~\cite{LHCb-PROC-2010-056}.
Decays of unstable particles are described by \evtgen~\cite{Lange:2001uf}, in which final-state radiation is generated using \photos~\cite{davidson2015photos}.
The interaction of the generated particles with the detector, and its response, are implemented using the \geant toolkit~\cite{Allison:2006ve, *Agostinelli:2002hh} as described in Ref.~\cite{LHCb-PROC-2011-006}.
The signal sample is generated with unpolarized \Lb decays and only a left-handed photon contribution.
The simulation is validated using the \LbpKJpsi and \LbLzJpsi, with \Jpsimumu,  control modes in data. 
The \LbpKJpsi mode is selected according to Ref.~\cite{LHCb-PAPER-2016-049}, while the \LbLzJpsi mode is selected with the strategy described for the signal mode below. 
The \Lb momentum distribution of all simulated samples with \Lb decays is corrected to data in two-dimensional intervals of \Lb transverse momentum and pseudorapidity, $\pt(\Lb)$ and $\eta(\Lb)$, using \LbpKJpsi background-subtracted data and simulated candidates. 

Offline signal candidates are reconstructed from the combination of a \Lz baryon and a high-energy photon candidate.
Opposite-charge good-quality track pairs, well separated from any PV, are combined to form \Lz candidates, where one track is consistent with the proton hypothesis and the other with the pion hypothesis.
Proton and pion candidates are required to have \pt larger than $800\mev$ and $300\mev$, respectively. 
The \Lz candidate must form a good-quality vertex that is well separated from the nearest PV and have a mass in the range $1110$--$1122\,\mev$.
Only \Lz candidates that decay in the central part of the vertex detector ($z < 270 \mm$) and have a \pt larger than $1\gev$ are retained for further study. 
Photons, reconstructed from clusters in the ECAL, must have $\et > 3\gev$ and be inconsistent with the extrapolation of reconstructed tracks to the ECAL.
Since the \Lb decay vertex cannot be reconstructed in this mode, the photon is assumed to originate from the $pp$ interaction region. 
The sum of the \Lz \pt and the photon \et must be larger than $5\gev$.
The \Lb candidate must have a transverse momentum above $4\,\gev$ and a mass within $900 \mev$ of the known \Lb mass~\cite{PDG2020}.
The distance of closest approach between the \Lb and \Lz trajectories is required to be small; the \Lb trajectory is assumed to originate from the PV closest to the \Lz trajectory

A boosted decision tree (BDT)~\cite{Breiman}, implementing the XGBoost algorithm~\cite{Chen:2016:XST:2939672.2939785}, used within 
the Scikit-learn library~\cite{Scikit-learn-paper}, separates signal from combinatorial background, which is formed by combinations of a real \Lz baryon with a random photon. The BDT classifier is trained on a signal sample of \LbLzGam simulated events and a background sample from data candidates with a mass larger than $6.1\gev$. 
The same variables of Ref.~\cite{LHCb-PAPER-2019-010}, with the addition of the photon pseudorapidity, are used as input for the classifier.
To maximize sensitivity to the photon polarization, the BDT threshold is selected to reject $99\%$ of background and retain around $50\%$ of signal candidates.

\begin{figure}
	\centering
	\includegraphics[width=0.45\linewidth]{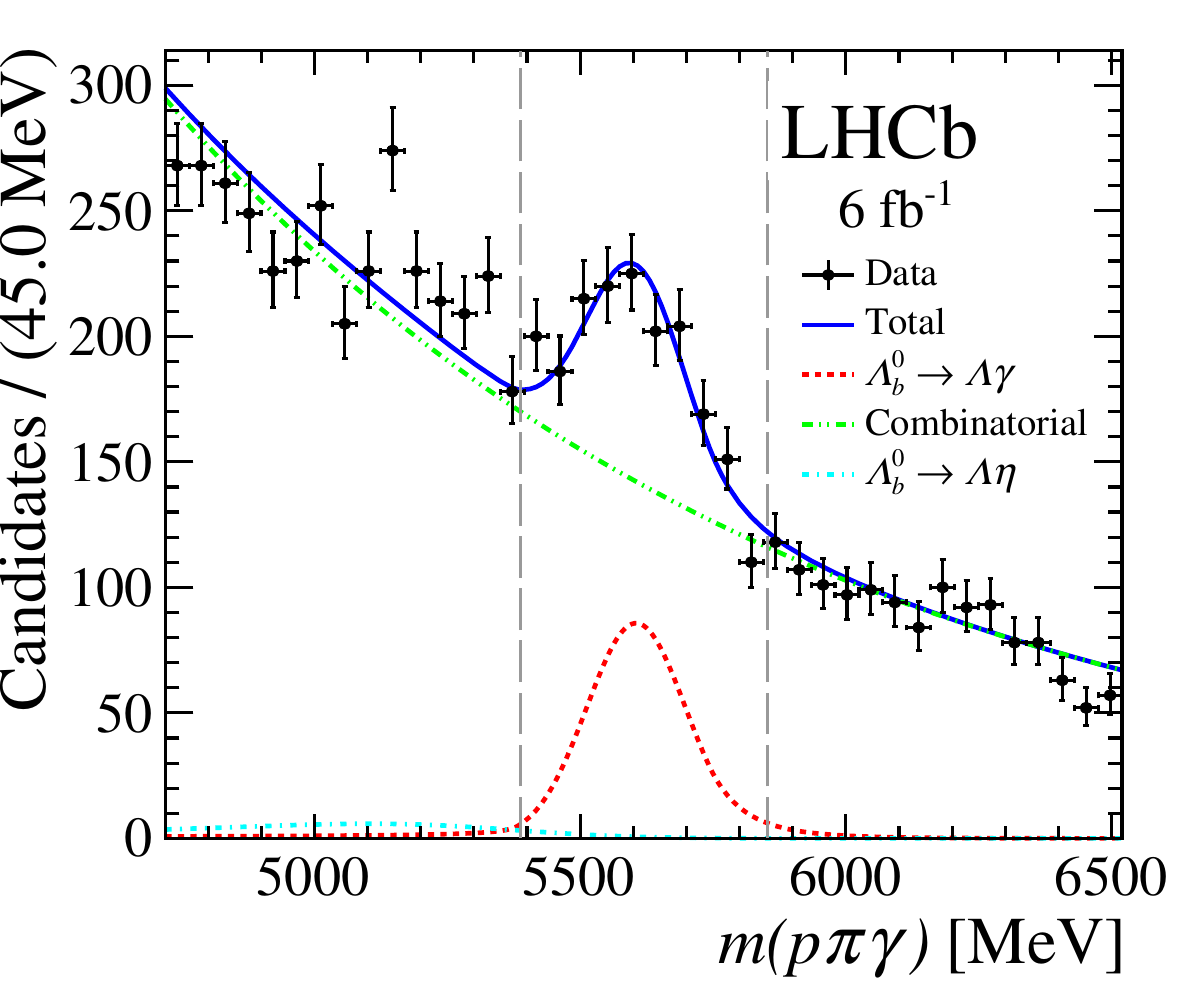}
	\includegraphics[width=0.45\linewidth]{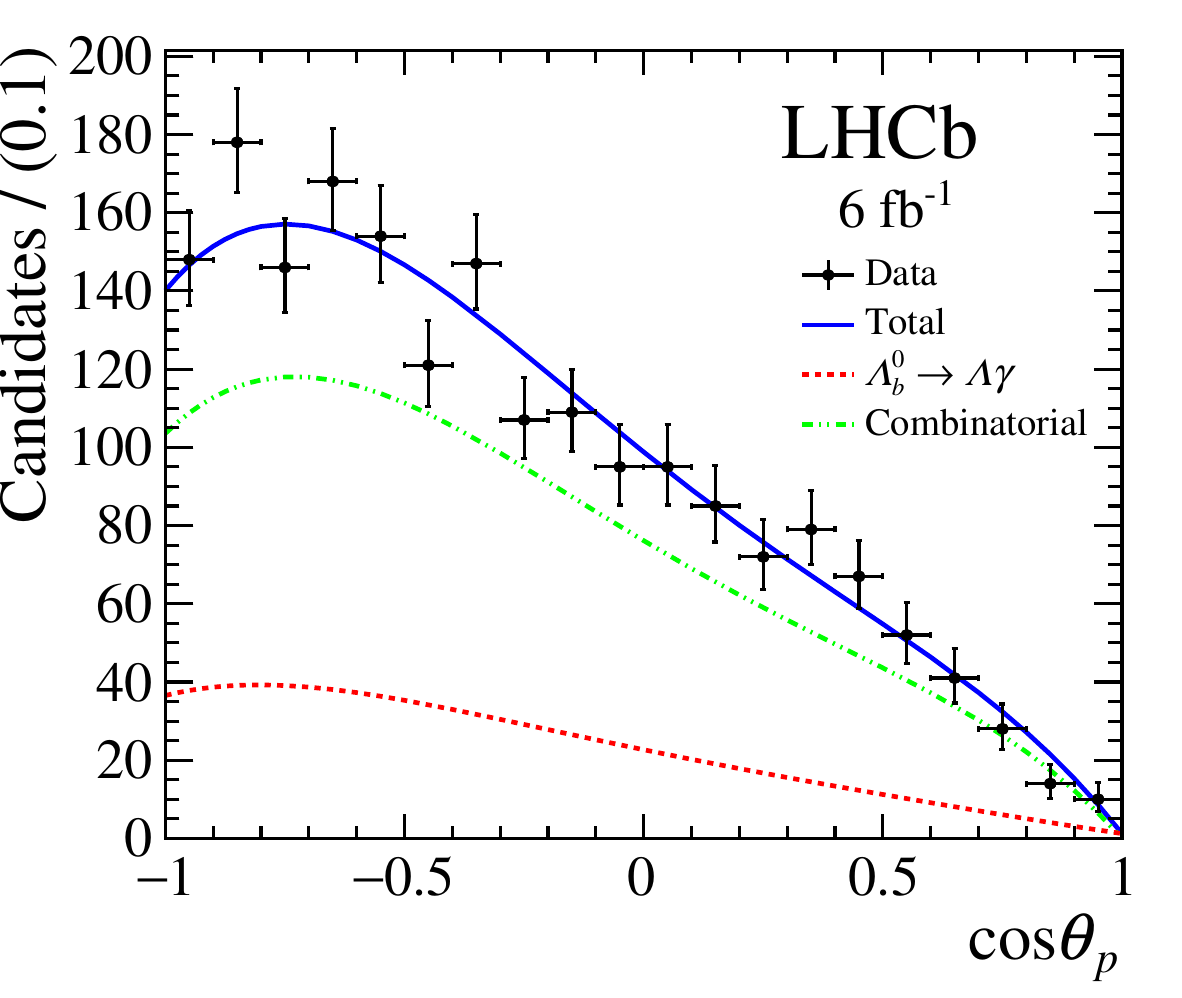}
	\caption{Distribution of (left) mass,  $m(\proton\pi\g)$, and (right) \ctp for \LbLzGam candidates. The latter shows the subset of candidates in the signal region $m(\proton\pi\g) \in [5387.1, 5852.1]\mev$, delimited by dashed vertical lines in the former.
		The results of the fits are overlaid as well as the 
		individual contributions. 
		\label{fig:fit}}
\end{figure}

The mass distribution of the selected candidates is shown in Fig.~\ref{fig:fit}.  A fit to this mass distribution is used to determine the signal and background yields.
The \LbLzGam signal component is modeled with a 
Crystal Ball~\cite{Skwarnicki:1986xj} probability density function (PDF), with power-law tails above and below the \Lb mass. 
The tail parameters and the width of the signal peak are fixed to values determined from simulation while the mean is allowed to vary freely in the fit to data.
A comprehensive set of background contributions is investigated but only two are found to be significant~\cite{LHCb-PAPER-2019-010}.
The dominant contribution is from combinatorial background
which is modeled by an exponential PDF with a freely varying slope.
A small contamination from \LbLzEta decays with \Etagg, where one of the photons is not reconstructed, is also expected and is described with the shape determined from simulation.
The signal and combinatorial background yields vary freely in the fit to data,
while the yield of \LbLzEta is constrained to its expected value, obtained from its measured branching fraction~\cite{PDG2020} and reconstruction and selection efficiencies determined from simulation.

The signal and background yields are obtained from an extended unbinned maximum likelihood fit to data, as shown in Fig.~\ref{fig:fit} (left).
In the signal region, $[5387.1, 5852.1]\mev$, the signal yield is $440\pm40$, with $1460\pm23$ and $10\pm4$ combinatorial and \LbLzEta decays, respectively.

The photon polarization is measured with an unbinned maximum likelihood fit to the \ctp distribution for events in the signal region, as shown in Fig.~\ref{fig:fit} (right).
The signal is described by the PDF in Eq.~\eqref{eq:Lb_AngDist} multiplied by an acceptance function, which accounts for the effect of the detector geometry, reconstruction and selection requirements.
The photon polarization parameter, \ag, is allowed to vary freely in the fit to data, whereas the physical boundary $[-1, 1]$ on this parameter is imposed later using the Feldman-Cousins technique~\cite{FeldmanCousins}. The \Lz weak decay parameter, \aL, is fixed to the average of the values measured by BESIII in \Lz and \Lbar decays, $\aL=0.754\pm 0.004$~\cite{BESIII:2018cnd}.
The shape of the acceptance in \ctp is determined from simulation and is parametrized as a fourth order polynomial. It is dominated by the impact parameter and transverse momentum requirements on the proton and the pion. 
The correct description of the acceptance in simulation is validated using \LbLzJpsi decays, which are reconstructed without using information from the \jpsi vertex. This decay mode, with higher yield and purity than that of the signal mode, has the same final-state hadrons.
The combinatorial and \LbLzEta components are described together by a fourth-order polynomial, with the coefficients determined from a fit to the candidates in the \Lb mass control region $[4719.6, 5382.1]\cup[5857.1, 6519.6]\mev$. The values of these coefficients are found to be consistent for candidates below and above the signal region.
The ratio between signal and background decays is constrained by the yields obtained from the mass fit.
The fit stability is validated with pseudoexperiments and the results were only examined after all
analysis procedures had been finalized in order to avoid experimenter's bias.
The result of the fit to data candidates is
$\ag = 0.82 \pm 0.23,$ where the uncertainty is obtained from the Hessian matrix. 

\begin{figure}
	\centering
	\includegraphics[width=0.45\linewidth]{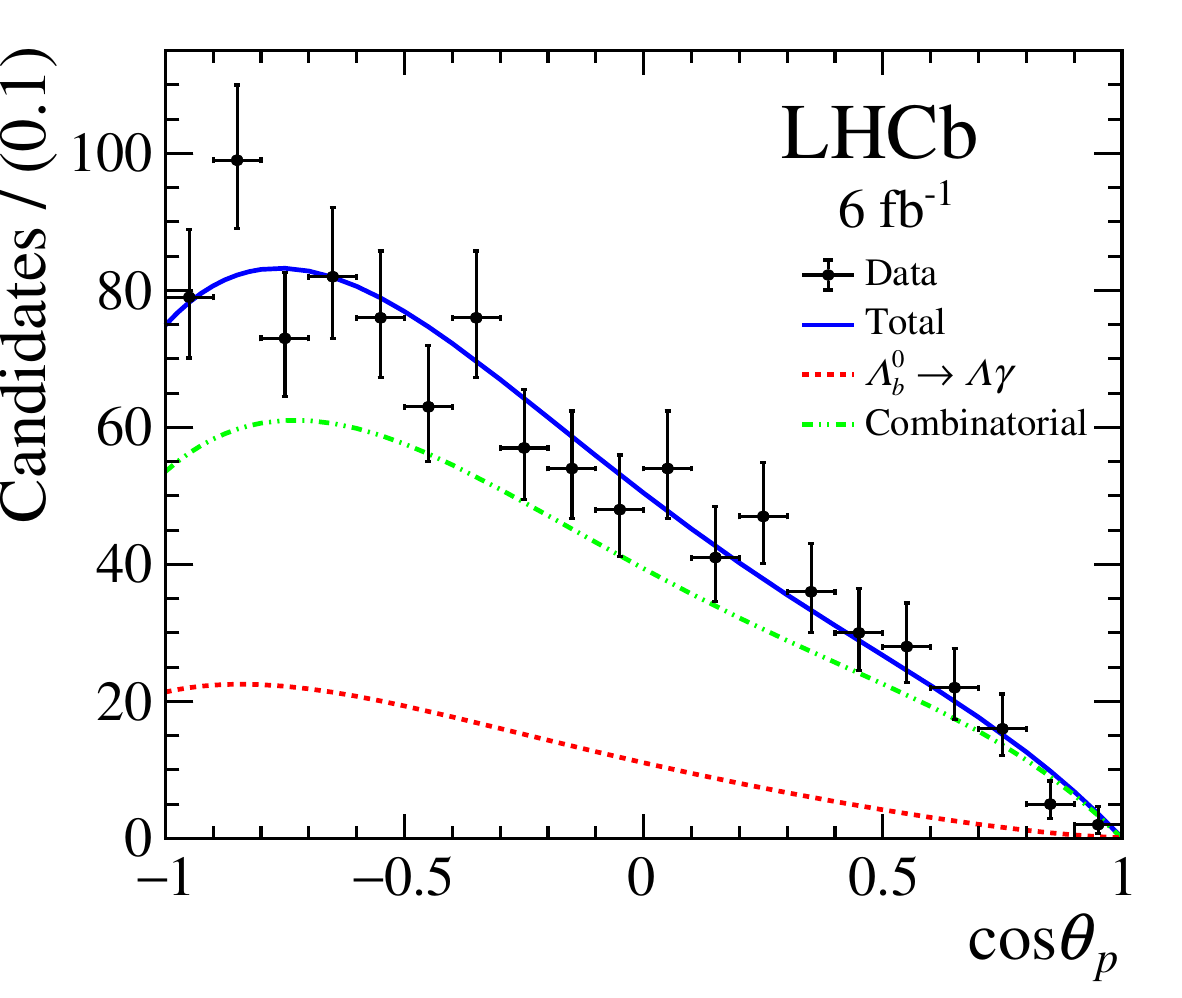}
	\includegraphics[width=0.45\linewidth]{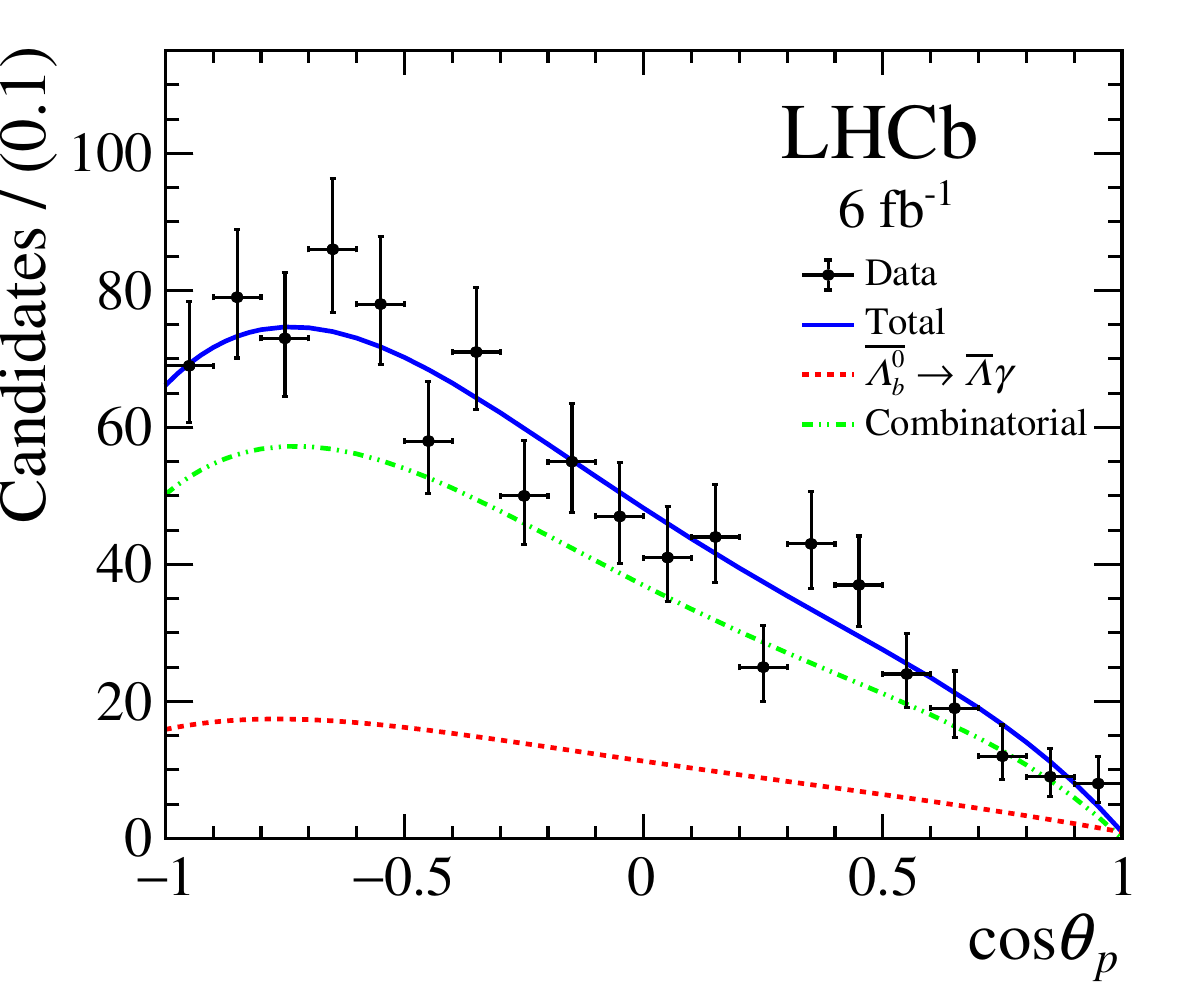}
	\caption{Distribution of \ctp for selected (left) \LbLzGam and (right) \LbbarLzbarGam candidates as well as the
		individual contributions.
		\label{fig:fit_cp}}
\end{figure}

The data are separated into \LbLzGam and \LbbarLzbarGam by the charge of the final-state pion and potential \CP breaking effects on the photon polarization are studied.
The photon polarization values in \LbLzGam and \LbbarLzbarGam decays, denoted as \agm and \agp, are expected to have the same absolute value with opposite sign in the absence of \CP violation.
Both values are measured independently for \Lb and \Lbbar decays following the same strategy as described for the combined sample.
The shape of the mass distribution for \Lb and \Lbbar decays is studied in simulation and is compatible, for both signal and \LbLzEta decays, therefore the shape from the combined sample is used in the fit of the two data sets. 
In the signal region, $233 \pm 32$ \LbLzGam and $210 \pm 30$ \LbbarLzbarGam decays are found. These yields are affected by production and detection asymmetries~\cite{LHCb-PAPER-2021-016} and cannot be directly interpreted as a measurement of \CP-violation effects. 
The \Lz and \Lbar weak-decay parameters measured by BESIII~\cite{BESIII:2018cnd}, $\aLm = 0.750 \pm 0.009 \pm 0.004$ and $\aLp = -0.758 \pm 0.010 \pm 0.007$, are used in the angular fit for \LbLzGam and \LbbarLzbarGam, respectively.
Due to the limited sample size, the acceptance and background shapes are taken from the combined sample. 
The fit to data candidates yields $\agm = 1.26 \pm 0.42$ and $\agp = -0.55 \pm 0.32$, where the uncertainty is statistical only. The results of the angular fits are shown in Fig.~\ref{fig:fit_cp}.
The measured value of \agm is found to be affected by a $5\%$ bias and a $13\%$ over-coverage driven by boundary effects in this observable.

Systematic uncertainties associated with the determination of the signal and background yields in the mass fit, the modeling of the background shape and signal acceptance and the finite precision on the external parameter \aL are evaluated. These uncertainties are determined by fitting pseudosamples, generated by an alternative model, with the default model.
The limited data size from the mass control region used to determine the background angular shape dominates this uncertainty and is evaluated by varying the background shape parameters. The systematic uncertainty of the background parameterization is evaluated using a third and fifth degree polynomial. 
The acceptance uncertainty is evaluated by considering the limited size of the simulation, using alternative fifth and third degree parametrizations, and removing the \Lb kinematic corrections.
The mass fit uncertainty is determined with pseudoexperiments using an alternative model with a variable signal width.
Finally, the limited precision of the \Lz decay parameter introduces a small systematic uncertainty. The photon polarization, with all systematic uncertainties included, is $\ag = 0.82 \pm 0.23 \pm 0.13$.

For the \CP asymmetry measurement, the uncertainties described above are $100\%$ correlated and sum to 0.13 and 0.11 for \agm and \agp, respectively. 
Additional uncertainties arise from potential differences in the background shape and signal acceptance between \Lb and \Lbbar decays. Fits are performed with the default combined model to pseudosamples generated with alternative models obtained independently for \Lb and \Lbbar candidates, to estimate this effect.
These uncertainties are uncorrelated and are 0.15 and 0.12 for \agm and \agp, respectively, resulting in $\agm = 1.26 \pm 0.42 \pm 0.20$ and $\agp = -0.55 \pm 0.32 \pm 0.16$.

Confidence intervals on the photon polarization are set within the physical limits of $[-1,1]$, see Eq.~\ref{eq:photon_polarization}, using the Feldman-Cousins technique, the fit result, and the relation between the true and measured values obtained from pseudoexperiments, including both statistical and systematic uncertainties.
The photon polarization in \LbLzGam decays is found to be
\begin{equation*}
     \alpha_\gamma = 0.82^{\,+\,0.17}_{\,-\,0.26} \text{ (stat.)}^{\,+\,0.04}_{\,-\,0.13} \text{ (syst.)},
\end{equation*}
which is compatible with the SM prediction.
This result excludes two regions allowed by previous measurements for the Wilson coefficients of the effective Hamiltonian of the \btosgam transition (see the Supplemental material to this Letter for details). 

The Feldman-Cousins technique is also applied to the photon polarization measurements in \LbLzGam and \LbbarLzbarGam decays, which are found to be
\begin{eqnarray*}
    \agm &>& 0.56\; (0.44) \text{ at 90\% (95\%) CL},\\
    \agp &=& -0.56^{\,+\,0.36}_{\,-\,0.33}\text{ (stat.)}^{\,+\,0.16}_{\,-\,0.09} \text{ (syst.)},\nonumber
\end{eqnarray*}
consistent with \CP symmetry.


To summarize, the photon polarization in \btosgam transitions has been measured for the first time in \mbox{\bquark-baryon} decays exploiting the decay chain \LbLzGam, \Lzppi, which enables the determination of this observable from the angular distribution of the \Lz decay products.
The \ag result is compatible with the SM prediction at the level of one standard deviation and provides new constraints that are compatible with previous measurements of the photon polarization in \bquark-meson decays~\cite{LHCb-PAPER-2020-020,LHCb-PAPER-2019-015}.
Additionally, \CP-breaking effects on this observable are studied for the first time and the results are compatible with \CP symmetry. 
These results can be used to place constraints on the real and imaginary contributions of right-handed currents in BSM models.
The precision of the results is dominated by the statistical uncertainty; furthermore, the dominant systematic uncertainty, due to the angular shape of the background, can be reduced with more data. Consequently, future data that will be collected by the \lhcb experiment during the next decade will enable significant improvements of these results.

\section*{Acknowledgements}
%
%
\noindent We express our gratitude to our colleagues in the CERN
accelerator departments for the excellent performance of the LHC. We
thank the technical and administrative staff at the LHCb
institutes.
We acknowledge support from CERN and from the national agencies:
CAPES, CNPq, FAPERJ and FINEP (Brazil); 
MOST and NSFC (China); 
CNRS/IN2P3 (France); 
BMBF, DFG and MPG (Germany); 
INFN (Italy); 
NWO (Netherlands); 
MNiSW and NCN (Poland); 
MEN/IFA (Romania); 
MSHE (Russia); 
MICINN (Spain); 
SNSF and SER (Switzerland); 
NASU (Ukraine); 
STFC (United Kingdom); 
DOE NP and NSF (USA).
We acknowledge the computing resources that are provided by CERN, IN2P3
(France), KIT and DESY (Germany), INFN (Italy), SURF (Netherlands),
PIC (Spain), GridPP (United Kingdom), RRCKI and Yandex
LLC (Russia), CSCS (Switzerland), IFIN-HH (Romania), CBPF (Brazil),
PL-GRID (Poland) and NERSC (USA).
We are indebted to the communities behind the multiple open-source
software packages on which we depend.
Individual groups or members have received support from
ARC and ARDC (Australia);
AvH Foundation (Germany);
EPLANET, Marie Sk\l{}odowska-Curie Actions and ERC (European Union);
A*MIDEX, ANR, IPhU and Labex P2IO, and R\'{e}gion Auvergne-Rh\^{o}ne-Alpes (France);
Key Research Program of Frontier Sciences of CAS, CAS PIFI, CAS CCEPP, 
Fundamental Research Funds for the Central Universities, 
and Sci. \& Tech. Program of Guangzhou (China);
RFBR, RSF and Yandex LLC (Russia);
GVA, XuntaGal and GENCAT (Spain);
the Leverhulme Trust, the Royal Society
 and UKRI (United Kingdom).




\clearpage

\section{Supplemental material for PRL}
\label{sec:Supplemental-App}
\begin{figure}[!htb]
  \begin{center}
    \includegraphics[width=0.85\textwidth]{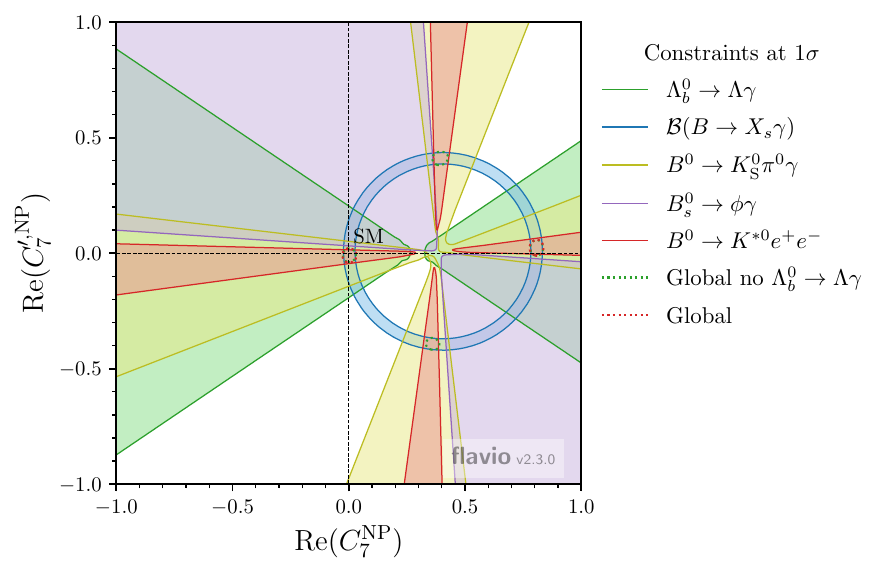}
    \includegraphics[width=0.85\textwidth]{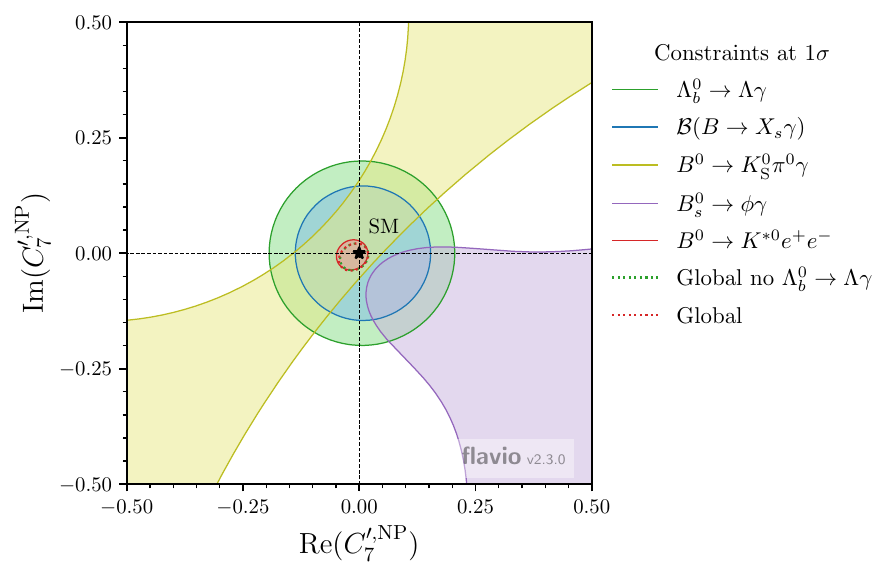}
  \end{center}
  \caption{
    \small 
    Constraints at the $1 \sigma$ level on the new physics contributions to the (top) left- and right-handed Wilson coefficients, \Ceff7 and \Cpeff7, and on the (bottom) real and imaginary parts of \Cpeff7. In the latter, the \Ceff7 coefficient is fixed to its SM value. 
    The measurements of the inclusive branching fraction, $\BF(\decay{B}{X_s \g})$, and the $\decay{\Bd}{\KS \piz \g}$ mixing-induced
\CP asymmetry by the Belle and BaBar experiments~\cite{Belle:2006pxp,BaBar:2008okc} are shown in blue and yellow,
respectively, the \BsPhiGam and \BdKstee measurements by the \lhcb experiment~\cite{LHCb-PAPER-2019-015,LHCb-PAPER-2020-020} in purple and red, respectively, and the \ag measurement presented
in this paper in green. The global fit is shown in dashed red (green) lines including (excluding) this measurement.}
  \label{fig:flavio}
\end{figure}

The measurement of \ag in this paper is used to place constraints on the electromagnetic dipole Wilson coefficients of the effective Hamiltonian of the \btosgam transition, namely \Ceff7 and \Cpeff7 .
The leading order expression of \ag in terms of Wilson coefficients is used within the \texttt{flavio} package~\cite{Straub:2018kue} to this end. 
This enables the experimental exclusion of two solutions with large real values of \Ceff7 and \Cpeff7, which, although phenomenologically unfavored, were experimentally allowed by all the previous measurements, as shown in Fig.~\ref{fig:flavio} (top).

\begin{figure}[!htb]
  \begin{center}
    \includegraphics[width=0.45\textwidth]{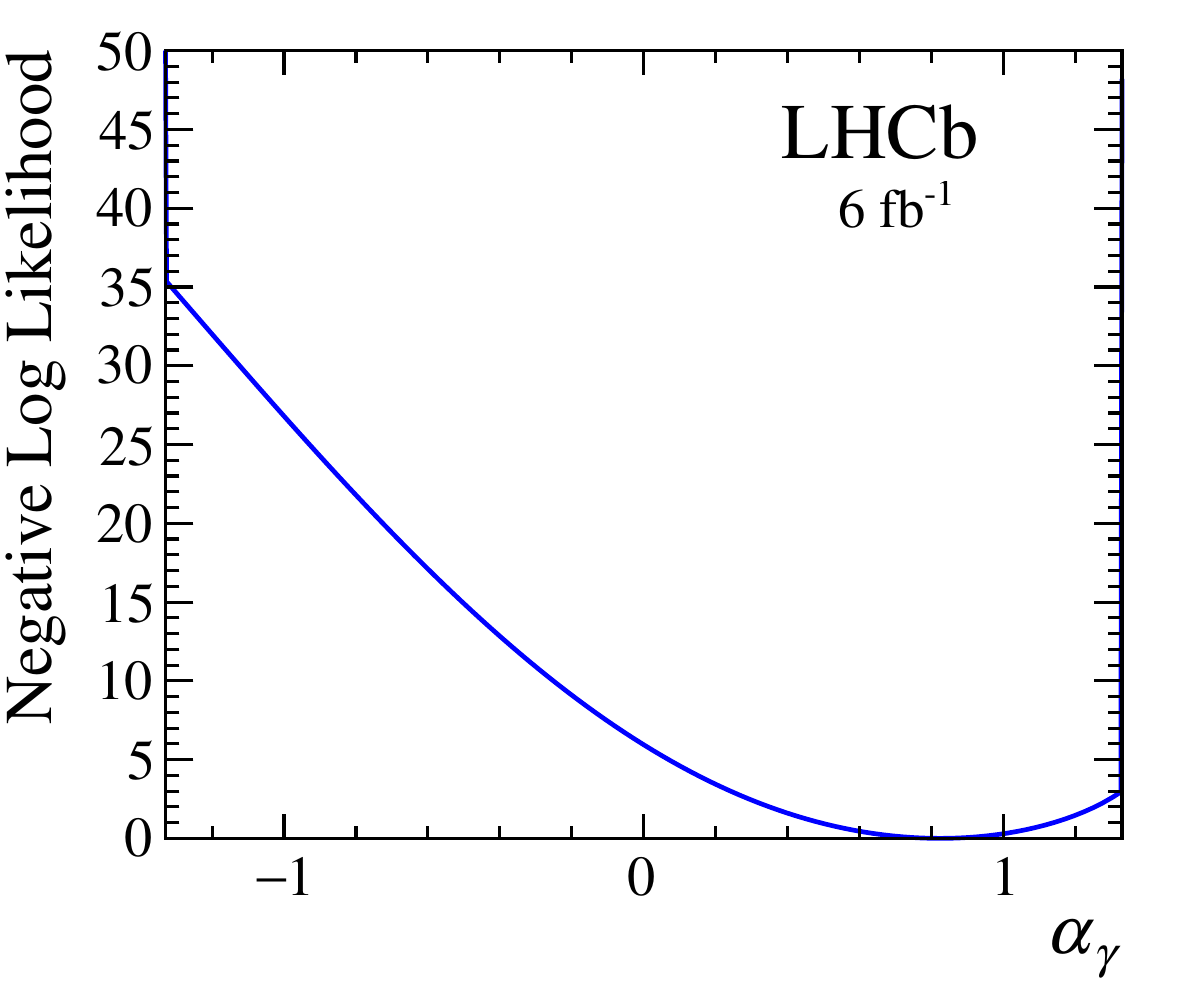}\\
    \includegraphics[width=0.45\textwidth]{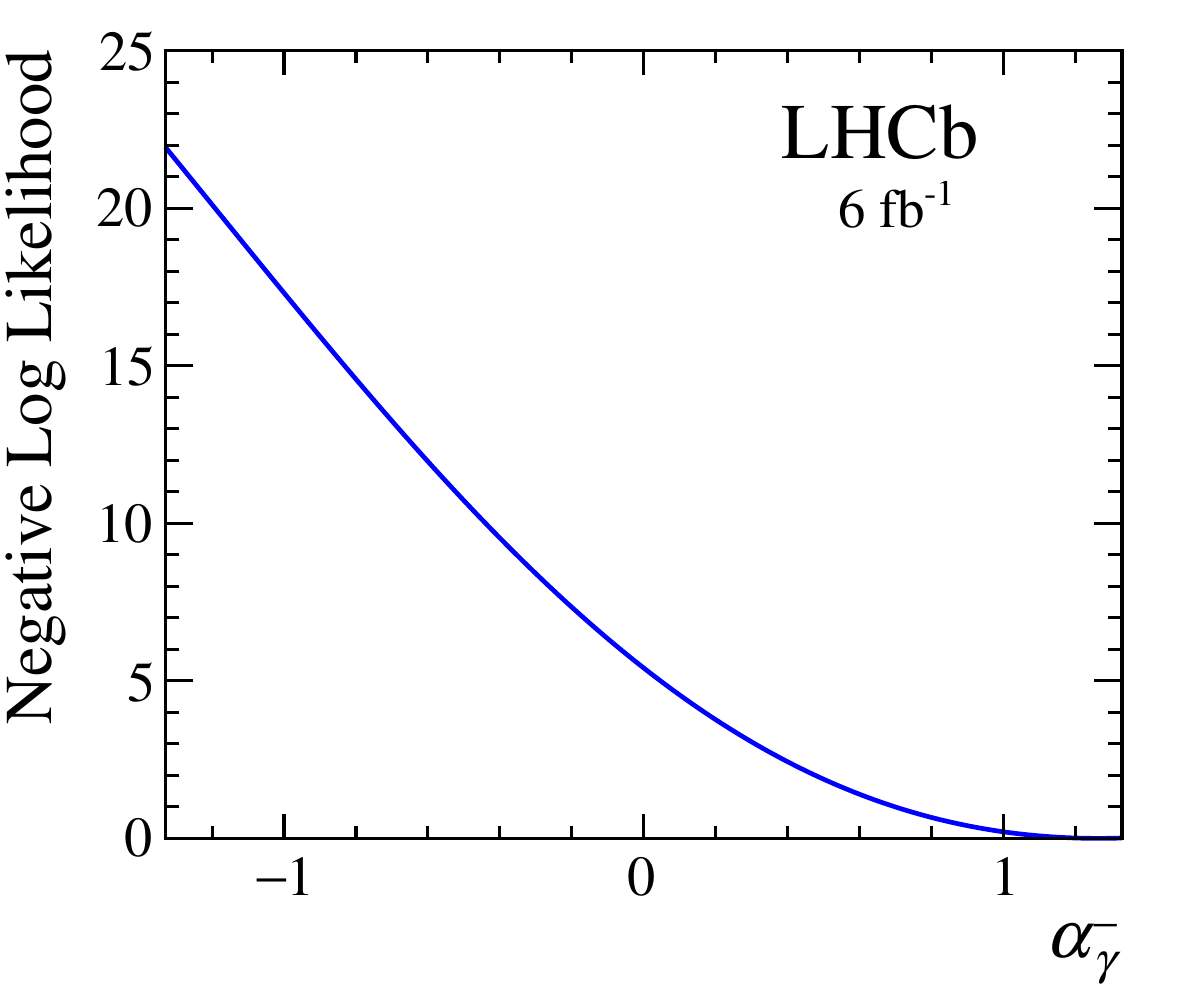}
    \includegraphics[width=0.45\textwidth]{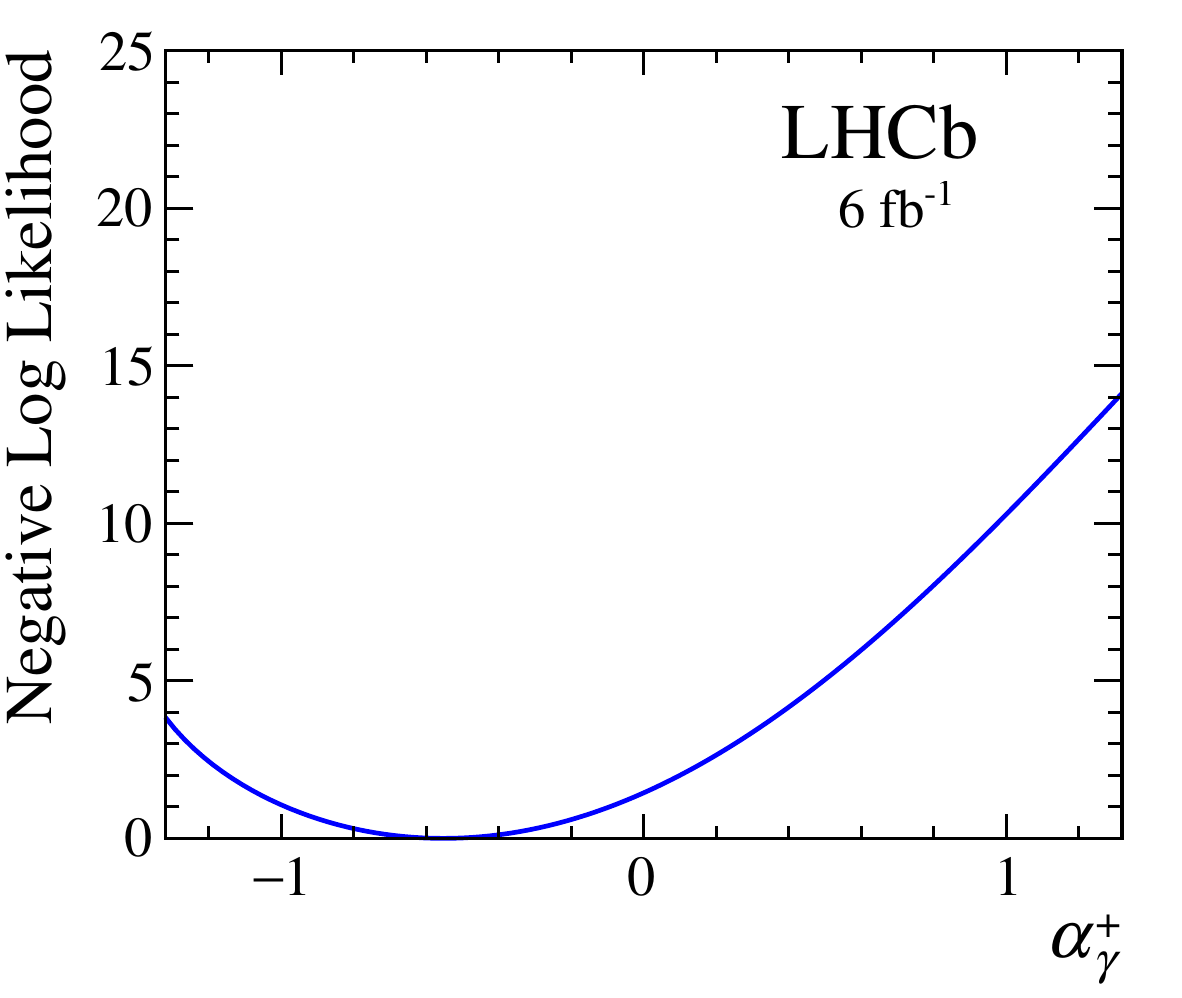} 
  \end{center}
  \caption{
    \small 
    Profile of the negative log likelihood along the (top) combined, (bottom-left) \LbLzGam and (bottom-right) \LbbarLzbarGam photon polarization values (\ag, \agm and \agp, respectively).}
  \label{fig:fc_cp}
\end{figure}

\clearpage

\addcontentsline{toc}{section}{References}
\bibliographystyle{LHCb}
\bibliography{main,standard,LHCb-PAPER,LHCb-CONF,LHCb-DP,LHCb-TDR}

\newpage
\centerline
{\large\bf LHCb collaboration}
\begin
{flushleft}
\small
R.~Aaij$^{32}$,
A.S.W.~Abdelmotteleb$^{56}$,
C.~Abell{\'a}n~Beteta$^{50}$,
F.~Abudin{\'e}n$^{56}$,
T.~Ackernley$^{60}$,
B.~Adeva$^{46}$,
M.~Adinolfi$^{54}$,
H.~Afsharnia$^{9}$,
C.~Agapopoulou$^{13}$,
C.A.~Aidala$^{87}$
% ...

{\footnotesize \it

% ...
$^{9}$Universit{\'e} Clermont Auvergne, CNRS/IN2P3, LPC, Clermont-Ferrand, France\\
% ...
$^{13}$LPNHE, Sorbonne Universit{\'e}, Paris Diderot Sorbonne Paris Cit{\'e}, CNRS/IN2P3, Paris, France\\
% ...
$^{32}$Nikhef National Institute for Subatomic Physics, Amsterdam, Netherlands\\
% ...
$^{46}$Instituto Galego de F{\'\i}sica de Altas Enerx{\'\i}as (IGFAE), Universidade de Santiago de Compostela, Santiago de Compostela, Spain\\
% ...
$^{50}$Physik-Institut, Universit{\"a}t Z{\"u}rich, Z{\"u}rich, Switzerland\\
% ...
$^{54}$H.H. Wills Physics Laboratory, University of Bristol, Bristol, United Kingdom\\
% ...
$^{56}$Department of Physics, University of Warwick, Coventry, United Kingdom\\
% ...
$^{60}$Oliver Lodge Laboratory, University of Liverpool, Liverpool, United Kingdom\\
% ...
$^{68}$Syracuse University, Syracuse, NY, United States\\
% ...
$^{87}$University of Michigan, Ann Arbor, United States, associated to $^{68}$\\
\bigskip
% ...
}
\end{flushleft}

\end{document}